\documentclass{article}
\usepackage[a4paper, left=3cm, right=2.5cm, top=3cm, bottom=3cm]{geometry}
\usepackage[utf8]{inputenc}
\usepackage{etex} 
\usepackage[utf8]{inputenc}
\usepackage[T1]{fontenc}
\usepackage{setspace}
\usepackage[sc]{mathpazo}            
\usepackage{tabularx}
\usepackage{longtable}
\usepackage{multirow}
\usepackage{colortbl}
\usepackage{booktabs}
\usepackage{array, arydshln}
\usepackage{amsmath}
\usepackage{framed}
\usepackage{enumerate}             
\usepackage{enumitem}
\usepackage{textcomp}                %
\usepackage{fancyhdr}
\usepackage{graphics}
\usepackage[tableposition=top, font=small]{caption}
\usepackage{subcaption}
\usepackage{color}
\usepackage{pdfpages}
\usepackage[format=plain, textfont={footnotesize},labelfont=bf]{caption}
\usepackage[pdftex, plainpages=false, hypertexnames=false, pdfpagelabels=true,
    hyperindex=true, linktocpage, pdfa=true]{hyperref}
\hypersetup{hypertexnames=true} 

\begin{document}
\title{Identification of social groups and waiting pedestrians at railway platforms using trajectory data}
\author{Mira Küpper$^1$\footnote{* Correspondence: mkuepper@uni-wuppertal.de} $\;$ and Armin Seyfried$^1$ $^2$}
\date{$^1$ School of Architecture and Civil Engineering, University of Wuppertal, Germany\\
    $^2$  Institute for Advanced Simulation, Forschungszentrum Jülich,  Germany\\[2ex]
}

%

\maketitle
\thispagestyle{empty}%
\hypersetup{pageanchor=false}
\setcounter{page}{1}
\thispagestyle{fancy}
\hrule
\begin{abstract}
To investigate the impact of social groups on waiting behaviour of passengers at railway platforms a method to identify social groups through the monitoring of distances between pedestrians and the stability of those distances over time is introduced. The method allows the recognition of groups using trajectories only and thus opens up the possibility of studying crowds in public places without constrains caused by privacy protection issues. Trajectories from a railway platform in Switzerland were used to analyse the waiting behaviour of passengers in dependence of waiting time as well as the size of social groups. The analysis of the trajectories shows that the portion of passengers travelling in groups reaches up to 10\% during the week and increases to 20 \% on the weekends. 60\% of the groups were pairs, larger groups were less frequent. With increasing group size, the mean speed of the members decreases. Individuals and pairs often choose waiting spots at the sides of the stairs and in vicinity of obstacles, while larger groups wait close to the platform entries. The results indicate that passengers choose waiting places according to the following criteria and ranking: shortest ways, direction of the next intended action, undisturbed places and ensured communication. While individual passengers often wait in places where they are undisturbed and do not hinder others, the dominating comfort criterion for groups is to ensure communication. The results regarding space requirements of waiting passengers could be used for different applications. E.g. to enhance the level of service concept assessing the comfort of different types of users, to avoid temporary bottlenecks to improve the boarding and alighting process or to increase the robustness of the performance of railway platforms during peak loads by optimising the pedestrian distribution.
\end{abstract}
\hrule
\section{Introduction}
The movement of pedestrians is studied in different situations, e.g. evacuations of buildings or large-scale events, and was analysed in both field and laboratory experiments; for an overview see \cite{predtetschenski1971, boltes2018, chraibi2018} and the conference series \cite{ped2018, tgf2020}. Most previous studies focused on characteristics of pedestrian flows and the parameters that influence their movement. The key concepts for evaluating a facility regarding pedestrian traffic include the fundamental diagram and the Level of Service (LOS) concept \cite{Weidmann1993, Buchmueller2006}.  The fundamental diagram is used to estimate the capacity of pedestrian facilities and whether heavy congestion occur. The Level of Service concept allows to rate the comfort at a certain density. While those concepts provide information on the comfort pedestrians feel and whether the pedestrian load can lead to dangerous situations, they were designed for environments in which movements occur. Waiting pedestrians were solely considered while standing in queues, as for example in the LOS presented in \cite{Fruin1971}; but obviously the interaction between moving and waiting and therewith standing pedestrians could have a great influence on the dynamic and must be considered in dependence of the context. Larger social groups in particular can be an impactful obstacle in pedestrian flows. This is for example the case at train station platforms, where both moving and waiting passengers are present simultaneously. The first part of this introduction focuses on research on waiting behaviour at train platforms, and the second part addresses aspects of the research on social groups.

Only limited research has been published to date about pedestrian waiting behaviour and the factors that influence how and where pedestrians wait at railway platforms and how these waiting positions influence the pedestrians moving on the platform. Previous studies on this topic revealed that pedestrians tend to cluster in the vicinity of the platform entries \cite{Pettersson2011, zhou2020, zhou2019, krstanoski2014} and around obstacles \cite{Bosina2015}.  Seating arrangements are frequently used as waiting place, beginning with the ones closest to the platform entries \cite{Pettersson2011}. The ticket machines were found to lead to crowding and congestion \cite{Davidich2013, Lam1999}, especially if placed close to stairs. Even if the stopping positions of trains are indicated on the overhead signals, passengers tend not walk to the farther ends of the platform as they do not trust the information or are unaware of its existence \cite{Pettersson2011}. Reference \cite{ingvardson2018} state that a confusing station layout leads to longer passenger waiting times, as pedestrians tend to arrive to such stations earlier in order to ensure they find their way. Moreover, the type of passengers using the train station platforms has an impact on the distribution and waiting behaviour. Depending on the purpose of the journey, passengers carry different amounts of luggage and possess varying degrees of familiarity with the environment. Passengers carrying luggage e.g., become important when the vertical or horizontal gaps between platform and train are larger \cite{lee2007}, as those will increase the boarding time. Commuters often develop individual strategies \cite{Bosina2017b} to minimise travel times and therefore for example wait in places where the train car that provides the shortest way at the desired destination is expected to arrive. 
This literature review highlights the previous studies on waiting pedestrians in the context of railway platforms. Those studies made no differentiation whether the pedestrians were individual persons travelling alone or members of social groups. 
Such a distinction however is necessary in order to interpret the findings and to respect the characteristics of individuals and social groups. This differentiation can help to sharpen the findings obtained on passenger's waiting behaviour.

Instead of considering a pedestrian crowd as consisting solely of a certain number of separate individuals who have no social relation, a crowd is rather to be understood as a gathering of individuals and small groups that are at the same place at the same time \cite{turner1987, drury2020, templeton2020}. The dynamics of inter-group behaviour are proposed as social identity theory and self-categorisation theory by \cite{tajfel1978} and \cite{turner2010}. The effect of group behaviour on pedestrian movements has become a growing research area.     
The following  will present the main findings of previous studies, which reveal differences between (moving) groups and individual persons in public environments. Depending on the group size and the density conditions social groups are expected to walk in specific manners: small groups of two to three members tend to walk side by side in low density environments \cite{moussaid2010} and form lines perpendicular to the groups walking direction, causing such groups to occupy a large area. With increasing density and therewith limited available space, groups adapt their walking behaviour and move in ``V" or ``U"-like formations \cite{moussaid2010, schultz2014, zanlungo2014}. Usually the central pedestrian in those configurations walks in the rear, ensuring the groups communication. Large groups split up into smaller subgroups, since communication with all group members becomes impossible \cite{moussaid2010}. Groups are slower than individuals and with increasing group size the velocity of the group members reduces; this was observed regardless of the density \cite{zanlungo2014b, zanlungo2014,zanlungo2015}. However, in high density conditions the velocity differences between members of social groups and individual persons become smaller, as groups give up their social interaction in favour of collision avoidance and start walking in single file \cite{zanlungo2015}. 
\cite{james1953} and \cite{coleman1961} analysed group sizes of free-forming small groups and found that each group size is less frequent than the next smaller group size.

In field observations social groups can be identified by the relation of their members. This relation is indicated by communication that is composed of oral and non-verbal elements such as gestures, body language and eye contact. 
Recent technical achievements in data collection (c.f. \cite{pouw2021,corbetta2014}) enabled the collection of large trajectory data sets, which prompted the development of methods to analyse the movement of social groups. For example \cite{ moussaid2010,zanlungo2014,zanlungo2014b,  zanlungo2015, yucel2013, yucel2019, brscic2014} use video recordings and trajectory data from public spaces to analyse and develop dynamical models for the movement of pedestrians in groups. The combination of video recordings and trajectory data offers the possibility to generate an annotated data set, in which information extracted from the videos, e.g., the visual identification of socially related pedestrians, can be transferred to the trajectories. Such an approach enables the analysis of the data of known social group members with respect to interpersonal distances, motion direction or angles between the group members velocity vectors.  However, all these studies focus on pedestrians that are walking. 
A method using trajectory data applicable for waiting pedestrians was proposed by \cite{pouw2020}. Social groups at train station platforms were identified based on their space and time relation. Pedestrians who showed a pairwise distance of below 1.5 m for 90\%  and a distance below 1 m for 40 \% of the time they spend at the platform were considered to be a social group (see also section \nameref{sec:method}). The study was performed with data collected in the first phase of the COVID-19 pandemic in 2020 and analysed with respect to contact tracing and distancing rules. 
\\
Up to 70 \% of pedestrians moving in urban environments can be assigned to social groups \cite{moussaid2010, aveni1977}, during events (such as sport events or public celebrations) the portion can be even higher \cite{schultz2014}. It than follows that the presence of social groups impacts the dynamics of pedestrian flows. 
Simulations indicate that large groups behave as moving obstacles \cite{reuter2012}. \cite{templeton2018} found that social groups walk slower, further and maintained closer proximity than non-group members. The presence of social groups influenced other pedestrians to walk faster and at a greater distance (even in counter-flow) in order to avoid moving inside the group. The characteristics of movement and walking configurations of social groups were also found to impact the evacuation processes. \cite{vonKruechten2017} reports on  a positive effect on the evacuation time when groups are present, as self-ordering processes were observed in the crowd at the exits. However, \cite{bode2015} performed egress experiments in which the presence of groups resulted in longer egress times, as members of social groups took longer to respond and move in the direction of the exits. 
\\
The findings of the studies presented above highlight the influence of social groups on the dynamic of crowds. It is expected that this also applies in the context of railway stations where both moving and waiting passengers are present. It is therefore of great interest to examine the influence of social groups on the capacity of pedestrian facilities such as railway platforms as well as in bottleneck situations like in the boarding and alighting process. Moreover, pedestrians that are members of social groups are expected to use the available space differently.
A current application of the results of such an analysis is the detection of offenders against the social distancing rules during the COVID-19 pandemic. While members of social groups, e.g., families, are allowed to have close contact, strangers are obliged to keep a distance from one another in order to reduce the risk of infection. To identify situations or regions in which the mandatory distance is not kept, the identification of social groups and individuals is essential.
\\
This paper seeks to fill the gap between existing studies on pedestrian waiting behaviour and research on social groups. Since only limited research has been published concerning the detection and analysis of characteristics of non-walking  social groups, this article presents a method to identify social groups at train station platforms, where both moving and waiting / standing behaviour is present. Based on the proposed method of group detection, the waiting behaviour of pedestrians at train station platforms is analysed. The use of trajectory data, in contrast to video recordings, ensures a privacy conserving methodology. The method is applied to data from a railway platform in Switzerland. The portions of pedestrians travelling in groups and the distribution of group sizes are analysed and the differences between members of social groups and individuals are discussed with respect to mean speed and choice of waiting places.

\section{Data Sources}
\begin{figure}[!h]
\includegraphics[width=\textwidth]{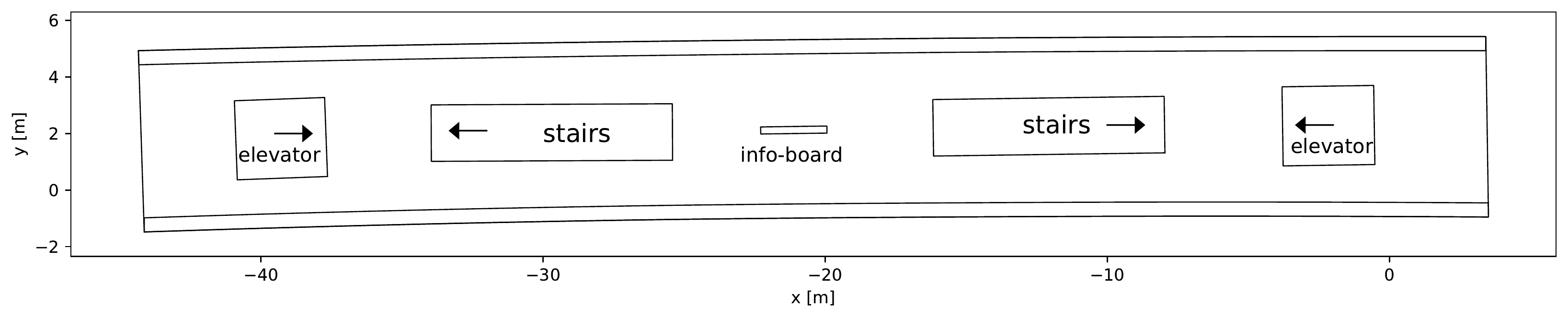}
\caption{ Spatial structure of the measurement area at the railway platform of Zürich Hardbrücke. Arrows indicate the movement direction for pedestrians entering the platform. The measurement area covers approximately 50 m of the platform.}
\label{fig_1}
\end{figure}

The tracking data used in this study was provided by the Swiss Federal Railways (SBB AG) and was collected at platform 2/3 of the station Zürich Hardbrücke, Switzerland. 
This train station platform is equipped with stereo sensors tracking the movement of pedestrians inside the area of observation with 10 frames per second. The data consists of an unique ID number for each pedestrian, a timestamp and the x and y coordinates of the pedestrian's position at the given timestamp.  As only the trajectories are recorded the data is fully privacy conserving and no information can be accessed that would allow to identify any individual pedestrian. The data was collected between $1^{st}$ and $28^{th}$ of February 2020 during the afternoon peak hours from 4 p.m. to 7 p.m. The data set thus consists of 8 weekend days and 20 workdays. The chosen time interval does not intersect with any measures introduced during the Covid-19 pandemic. The afternoon peak hours were selected with respect to comparability due to the fact that in these hours the passenger amount is usually high during both workdays and weekends and the most passengers travelling in social groups were expected. While the morning peak hours are often assigned to individual travel to e.g. work places, in the afternoon peak hours social activities are more likely. The observed area covers about 50 metres, see Fig~\ref{fig_1}. The platform  is constructed symmetrically with an information board in the central area and stairways and elevators to both sides. The direction of movement of passengers entering the platform is indicated with arrows at the stairs and elevators in Fig~\ref{fig_1}.

\subsection{Data Quality}
\label{subsec:Quality}
In order to assess the quality of the data, the starting and ending points of trajectories were checked for plausibility. Due to the setup of the measurement area, trajectories are expected to begin and end either at the platform entrances, the platform-train interfaces or at the sides of the measurement areas. Approximately 90 \% of all starting points of trajectories fulfil these requirements; the same applies to the ending points. However, in order to have a ``complete" trajectory, both beginning and ending are required to be at the expected positions. This is the case for about 75-80 \% of the trajectories.
Incomplete trajectories were caused by pedestrians being lost by the tracking system for one or more frames, as a new ID number is assigned upon re-detection (cf. \cite{vanHeuvel2017}). 
Nevertheless, those data are included in the analysis but the probability that the corresponding pedestrians are assigned as members of social groups is decreased.
Due to technical reasons, the tracking data is mirrored horizontally. 

\section{Methodology}
\label{sec:method}
In manual field observations or by watching video recordings social groups can be identified by their social interaction, observable through e.g. verbal communication, eye contact and gestures (cf. \cite{zanlungo2014}). This is not possible when working with trajectory data only. Nonetheless there are many research questions where a differentiation between social groups and individuals is crucial.

A method on group detection in railway environments was previously introduced by \cite{pouw2020} and used to perform contact tracing and analyse the distancing rules during the first phase of the Covid-19 pandemic. Their method uses a sparse graph in which the trajectories of the pedestrians as well as all events in which two pedestrians had a distance smaller than a predefined threshold of 2.5 m are memorized. Hence, the distances between each pair of pedestrians that is present simultaneously at the platform have to be calculated. The calculation of distances between N pedestrians scales with $N^2$ and is therefore very time-consuming for large N \cite{lohner2010}. While the distance calculation between all pedestrians is necessary for the analysis of Covid-19 distancing rules, it is not in the context of social group assessment. As members of social groups are expected to keep closer contact to each other than to non-group members, it is not necessary to calculate the distances between all pedestrians that are present at a given time. It is sufficient to determine the persons standing nearest to one another in every frame by applying Delaunay Triangulation which has a complexity ${\mathcal O}(N\cdot log(N))$ (cf. \cite{attali2003, abellanas1999}) and is therewith far more time efficient. 
Therefore, this paper proposes an adapted method to recognise social groups in trajectory data by analysing the distances and the stability of the distances between neighbouring persons which avoids the calculation of $N^2$ distances. 
 
Train station platforms are places where boarding and alighting passengers are present. Alighting passengers usually leave the train platforms in a straight path, which makes it almost impossible to determine socially related pedestrians, even if video recordings were available.
However, boarding passengers wait for a certain amount of time and can therefore be observed over longer time intervals. Hence, the identification of social groups is restricted to boarding passengers. The categorisation was performed by determining the start and end points of the trajectories: a boarding passengers trajectory starts at a platform entry and ends at a train door. A detailed discussion and analysis of that matter can be found in the authors' previous work, see \cite{kuepper2020}.  All trajectories that are shorter than 20 seconds were not included in the analysis, as this time interval was identified as minimal observation time needed for visual analysis (see section \nameref{sec:parameter}).

Since the group detection method identifies group members based on their distances from one another, the applicability is limited to low density environments. 

In consideration of the goal of determining members of social groups, which will be characterised by reasonably small distances, two different thresholds are defined for the distance between two pedestrians. A value of 1.5 metres was chosen as the maximum distance between persons for a contact ($d_{contact} \le 1.5$m) which was also established as social distancing threshold in numerous European countries during the Covid-19 pandemic and is the maximum  of the probability density of the pairwise distances in the data set. In order to regard the personal distance a value of 1 m is used ($d_{personal} \le 1$m) as pedestrians that are comfortable to be inside each others personal space over a longer period are most likely related in a social way.

Hence, $t_{contact}$ is determined as the number of frames $t$ for which holds
\begin{equation}
    \|\vec{X}_i(t) - \vec{X}_j(t) \| \le 1.5m = d_{contact}
\end{equation}
and $t_{personal}$ as the number of frames for which is
\begin{equation}
    \|\vec{X}_i(t) - \vec{X}_j(t) \| \le 1m = d_{personal}
\end{equation}
with $\vec{X}_i(t)$ being the position of pedestrian $i$ at time $t$, for pedestrian $j$ respectively.
In words, $t_{contact}$ translates to the number of frames in which the given pedestrians $i$ and $j$ are at a distance of 1.5 m or less from one another, and $t_{personal}$ as the number frames for which the distance is smaller than 1 metre. 
Since the two pedestrians $i$ and $j$ do not necessarily have to arrive and depart at the same time, the time in which both pedestrians $i$ and $j$ are inside the measurement area simultaneously, is calculated as
\begin{equation}
    t_{sim} = min\{t_{i,N}, t_{j,N}\} - max\{t_{i,0},t_{j,0}\}
\end{equation}
with $t_{i,0}$ representing the first frame  and $t_{i,N}$ the last frame in which pedestrian $i$ is inside the measurement area; for pedestrian $j$ respectively.

Following \cite{pouw2020} and \cite{yucel2019} the pedestrian pairs identified based on the small distances between them, will be checked for the following relations:  

\begin{equation}
    t_{contact} \ge \alpha \cdot t_{sim}
    \label{eq:t_cont}
\end{equation}
\begin{equation}
    t_{personal} \ge \beta \cdot t_{sim}
    \label{eq:t_pers}
\end{equation}
The values for $\alpha$ and $\beta$ are determined in a parameter study in the following section. If both Eq~(\ref{eq:t_cont}) and Eq~(\ref{eq:t_pers}) are fulfilled, the corresponding pedestrians $i$ and $j$ are considered to belong to the same social group. Groups with more than two members are detected by combination of pairs. 

\subsection{Parameter Study and Validation}
\label{sec:parameter}
To determine a suitable parameter set for $\alpha$ and $\beta$ and to validate the social groups found by the proposed method a ground truth of IDs that are members of social groups was established. To do so, the trajectory data of one example time interval of three hours was visualised as a video with JPSvis, which is the visualisation tool of the software JuPedSim \cite{Jupedsim2022}.

Two persons, who had no knowledge of the group detection, were asked to individually note all ID numbers of pedestrians, who they believe to be members of social groups. No specific instructions to the determination of groups were given, but both persons were asked for their strategy afterwards.  
The test persons identified group members based on simultaneous movements, similar waiting locations and close proximity over longer periods of time. It was monitored whether a certain person entered or left the area of observation along with others, or if the person stayed close to others during the time at the platform. A collective change of waiting positions was also used as indication of group affiliation. However, the visual recognition of groups based solely on trajectories is not a trivial procedure. In order to guarantee reliability, the results of the two test persons were compared. 
The first test person noted 154 IDs as members of social groups, the second 153 IDs. In total 146 IDs were listed by both testers, which means they agreed in $90.7\%$ of the cases.   
The IDs identified by both persons were used as ground truth for the parameter study. Therefore, all ID numbers that are not part of the 146 IDs found by both testers were considered to be individuals and in case one of those was found by the group detection method, it was assumed to be a false positive.

A parameter study was then performed to determine suitable values for $\alpha$ and $\beta$. Hence, suitable parameters are determined based on two constrains. The aim was to find a set of values for which a large number of members of social groups can be detected, however, the number of false positive detections should be zero, as those would correspond to pedestrians that are likely individuals but erroneously marked as group members. 

\begin{figure}[htbp]
\includegraphics[width=\textwidth]{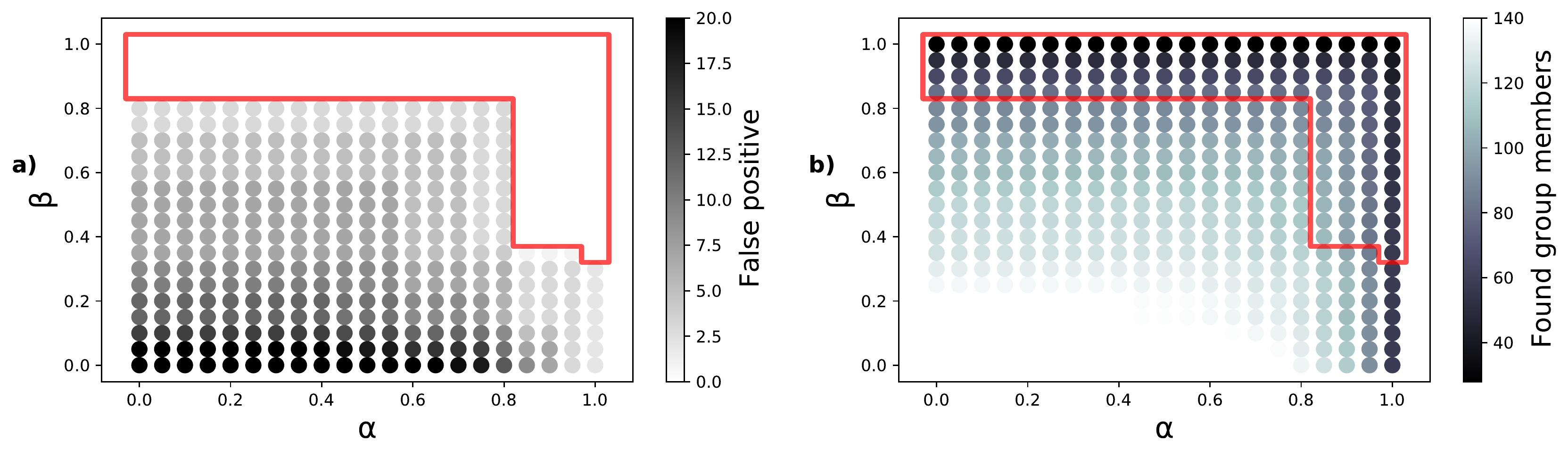}
\caption{ Parameter study. a) Number of false positive in the detection of group members by different values of $\alpha$ and $\beta$. White colours indicate that no false positives were found. b) Number of detected members of social groups. Light colours show high numbers of found members. The red area highlights the sets of parameters where the number of false positives in a) is zero. }
\label{fig_2}
\end{figure}

The values of $\alpha$ and $\beta$ were varied between 0 and 1 in steps of 0.05. For each set of values, the group detection was performed based on Eq~(\ref{eq:t_cont}) and Eq~(\ref{eq:t_pers}).  All IDs that were identified for a certain parameter set were than checked against the ground truth in order to determine any false positive.
The numbers of false positives increased with decreasing values for  $\alpha$ and $\beta$ (cf. darker colours in Fig~\ref{fig_2}a)). As the aim was to avoid false positives but to find a large number of group members, the numbers of identified group members for the corresponding parameters are illustrated in Fig~\ref{fig_2}b, with the red area marking the sets of parameters for which the number of false positives is zero. Therefore, the best parameter set will be identified as $\{\alpha, \beta\} = argmax(N(members), where\, N(false\, positive)=0)$, which corresponds to the set that produces the maximum number of found groups members within the red area.
From this it can be seen that the best results are achieved with  $\alpha = 0.85$ and $\beta=0.4$. In words, pedestrians are assumed to belong to a social group, if they have a distance smaller than 1.5 m  to at least one member of the group for 85\% of the time that they are simultaneously inside the measurement area and a distance smaller than 1 m for 40\% of that time. The work of \cite{pouw2020} proposed values of 90\% and 40\%, which can therewith be confirmed within 5\% by the performed parameter study.
This parameter combination allows for a maximum of 107 IDs to be correctly identified as group members without any false positives. This corresponds to about 73\% of the group members determined by the testers. Considering the overall data quality,  which exhibits about 75-80 \% of complete trajectories ( cf. section \nameref{subsec:Quality}), these results are satisfactory, as incomplete trajectories will interfere with the group detection method and prevent the correct assignment of group membership. 
\\
The method reaches its limits in crowded situations where higher densities are present over longer time intervals. In those cases, the close distance between pedestrians is not necessarily caused by social interaction but rather by limited available space. Due to the distance thresholds of 1 m and 1.5 m, the method can result in incorrect group assignment if the local density exceeds $0.5 - 1.0~ 1/ m^2$. If those densities remain over longer time intervals, crowding can be mistaken for social relation. The afternoon peak hours analysed in this study do not exhibit densities that exceed these threshold for longer time intervals. This will likely only occur in highly crowded situations, as e.g. in the context of public events.   
As the introduced method was developed for low density situations, no prediction can currently be made to what extent members of social groups preserve their close proximity in high density environments. It is expected that at least the distance thresholds and the parameters $\alpha$ and $\beta$ need to be adjusted in order to correctly determine social groups. It may be necessary to include additional criteria, like for example the simultaneous movement of nearby pedestrians.  However, the correlation of movement will not allow to expand the group detection to alighting passengers, since their walking paths inside the area of observation are generally similar and too short to allow assessment of group membership. In environments where pedestrians continuously move the correlation of group members can be expected to be higher than to unrelated pedestrians, even in increasing densities. In the context of train station platforms passengers spend most of their time waiting and therefore do not move, which will increase the correlation between (socially) unrelated neighbours in situations of limited available space (e.g. at a fully crowded platform).  

\subsection{Speed Calculation and Waiting}
The identification of waiting passengers on train platforms using trajectory data can be achieved by analysing their speed of movement or lack thereof.

The speed of a pedestrian at a given time is calculated as the movement of a pedestrian in a time interval $\Delta t$. With $\vec{x}_i(t)$ the location of the pedestrian at time $t$, the speed can be calculated as:

\begin{equation}
    v_i(t)= \frac{|\vec{x}_i(t + \Delta t'/2) - \vec{x}_i(t-\Delta t'/2)| }{\Delta t'}
    \label{eq:velocity}
\end{equation}
In this study $\Delta t = 50 $ frames, corresponding to 5 seconds, was used. To determine if a given pedestrian can be considered as waiting, a threshold of $v_i(t)< 0.4 $ m/s was applied. This threshold was picked as the local minimum of the velocity distribution of the data set, which shows two peaks: One peak at mean speeds of approximately 0.2 m/s; the other at 1.2 m/s.  The first peak mainly relates to boarding, the second to alighting passengers. The velocity distribution can be found in the author's previous work \cite{kuepper2020}.

\section{Results and Discussion}
\label{sec:discussion}
The group detection introduced in the previous section was used to analyse the differences in terms of numbers, mean speed and waiting positions between social groups and individuals at the train station platform in Zürich Hardbrücke (Switzerland). 
\begin{figure}[htbp]
\includegraphics[width=\textwidth]{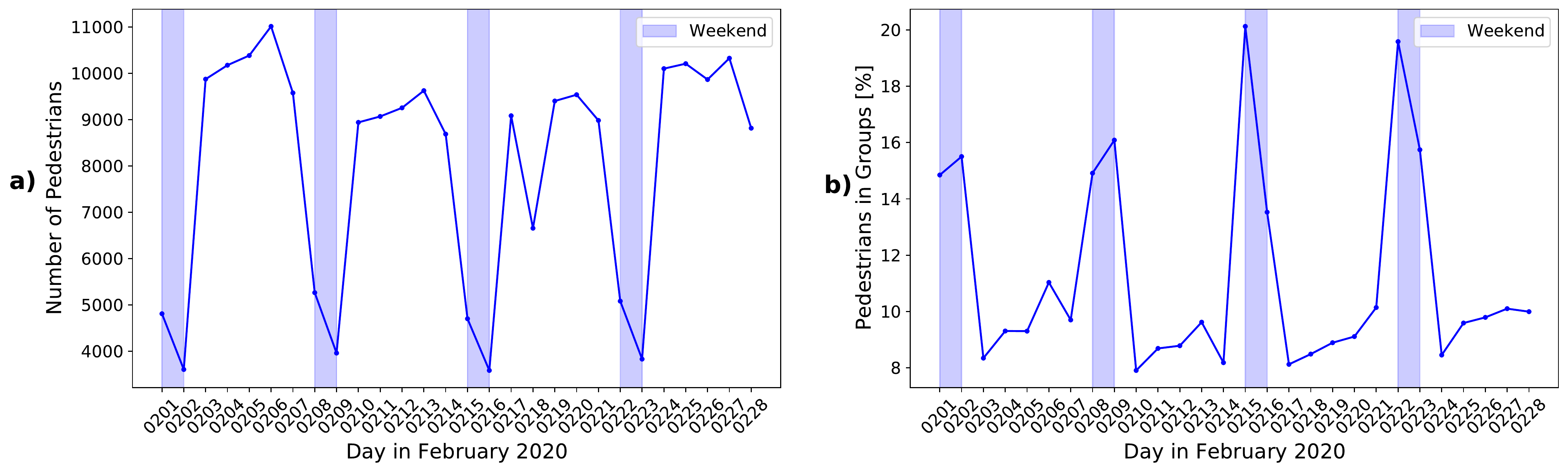}
\caption{ Number of passengers and members of social groups. a) Number of pedestrians during afternoon peak hours from 4 p.m. to 7 p.m. in February 2020 at Zürich Hardbrücke. b) Percentage of passengers who are members of a social group detected by the group detection method. Weekends are marked as blue bars.}
\label{fig_3}
\end{figure}
Due to recording errors, e.g., loss and re-detection of pedestrians, as well as to pedestrians leaving and re-entering the sensor area, the number of IDs in the sensor area is probably higher than the total number of pedestrians and might differ from passenger counts with other methods. However, in order to improve the readability of the following sections of this paper, the terms ``passenger" or ``pedestrian" are used as synonyms for ``ID".

On workdays between 9000 and 11000 passengers were detected in the observation area during the afternoon peak hours, cf. Fig~\ref{fig_3}a. On weekends the pedestrian load at the platform was lower and ranged between 4000 and 5000 passengers.  
The percentage of pedestrians that were identified as members of social groups by the method outlined in section \nameref{sec:method} ranged between 9 and 11 \% on workdays and rose to 14-20 \% on weekends, cf. Fig~\ref{fig_3}b. Weekday passenger traffic was dominated by individual passengers commuting to their work places, while during the weekends social activities played an important role. Due to the presence of data errors, e.g. incomplete trajectories, the actual number of members of social groups will be higher than the number of detected groups.

\begin{figure}[htbp]
\centering
\includegraphics[width=0.85\textwidth]{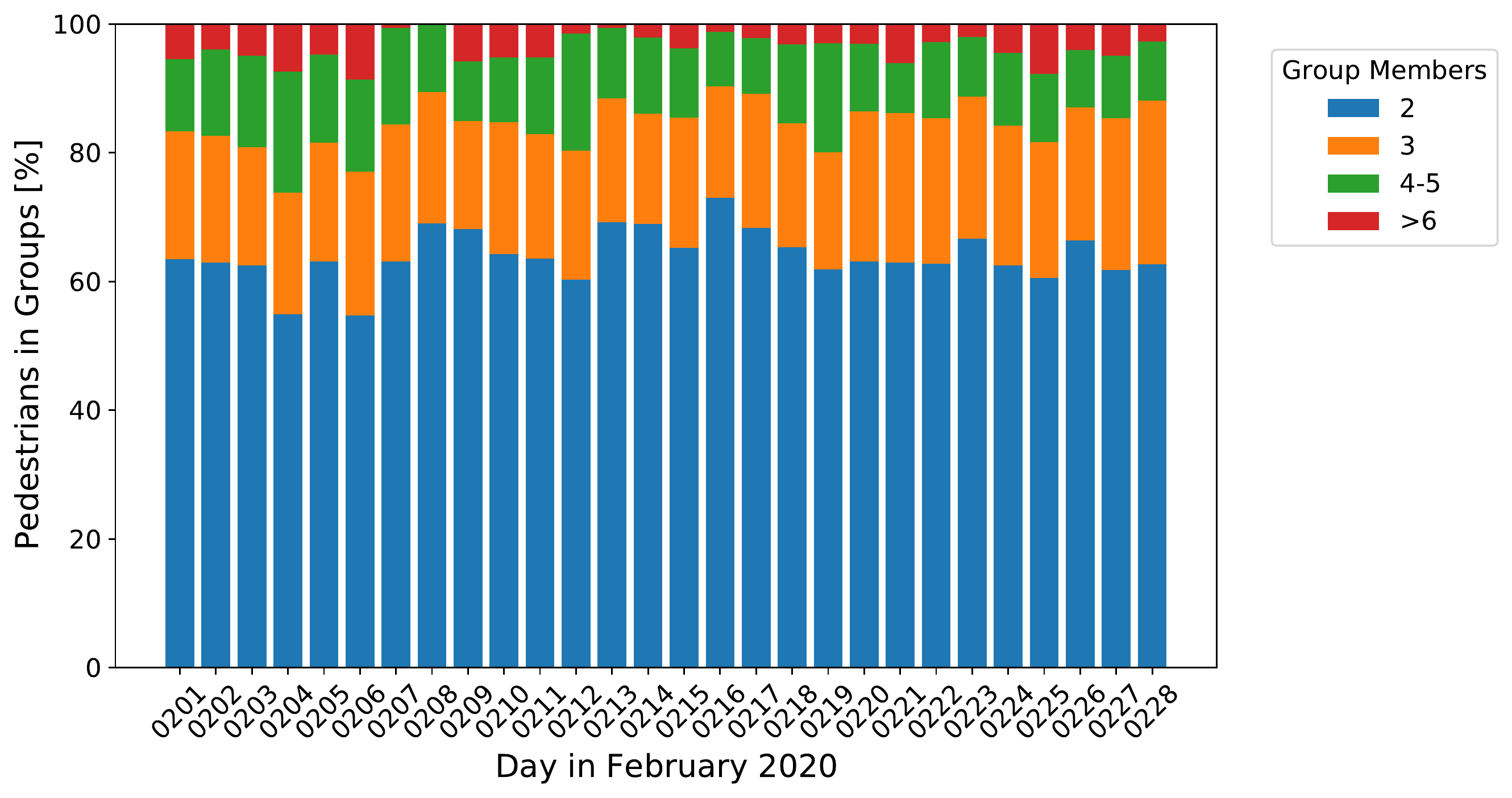}
\caption{ Distribution of group sizes as percentage of the number of passengers assigned to social groups. Each group size is less frequent than the next smaller group size. }
\label{fig_4}
\end{figure}

While the percentage of pedestrians who are members of social groups varies depending on workdays or weekends, the distribution of group sizes, as shown in Fig~\ref{fig_4} does not. In case of discontinuous trajectories (more than one ID number assigned to the trajectory of one pedestrian) the number of group members could be overestimated. In order to avoid this, the group size is determined as the maximal number of members present simultaneously. About 55\% to 70\% of all pedestrians assigned to social groups are members of pairs and approximately 20 \% of groups of three. Groups with four and more members are less frequent. In accordance with \cite{james1953}, smaller groups are more frequent than bigger groups and each size is less frequent than the next smaller group size. 
A similar analysis of distribution of group sizes was performed by \cite{pouw2020} for the time of the first phase of the Covid-19 pandemic in the Netherlands. Comparable to the results for the station in Switzerland more passengers are traveling in groups during the weekends, this seems to be independent of pandemic regulations. Groups with three or members are, however, less frequent in the work of \cite{pouw2020} which is also expected due to the pandemic restrictions and contact regulations considered. 

In order to determine whether or not individuals and members of social groups show different characteristics in terms of platform usage, the mean speed and waiting places were analysed with respect to group sizes. Since the distribution of group sizes is not affected by weekends and in order to increase the available data of social groups, the data set was accumulated over all days and analysed based on the group sizes. 
In total 15558 passengers were assigned to groups with two to three members, 1602 to groups with four to five members and 359 passengers to groups with six or more members.

\begin{figure}[htbp]
\centering
\includegraphics[width=0.7\textwidth]{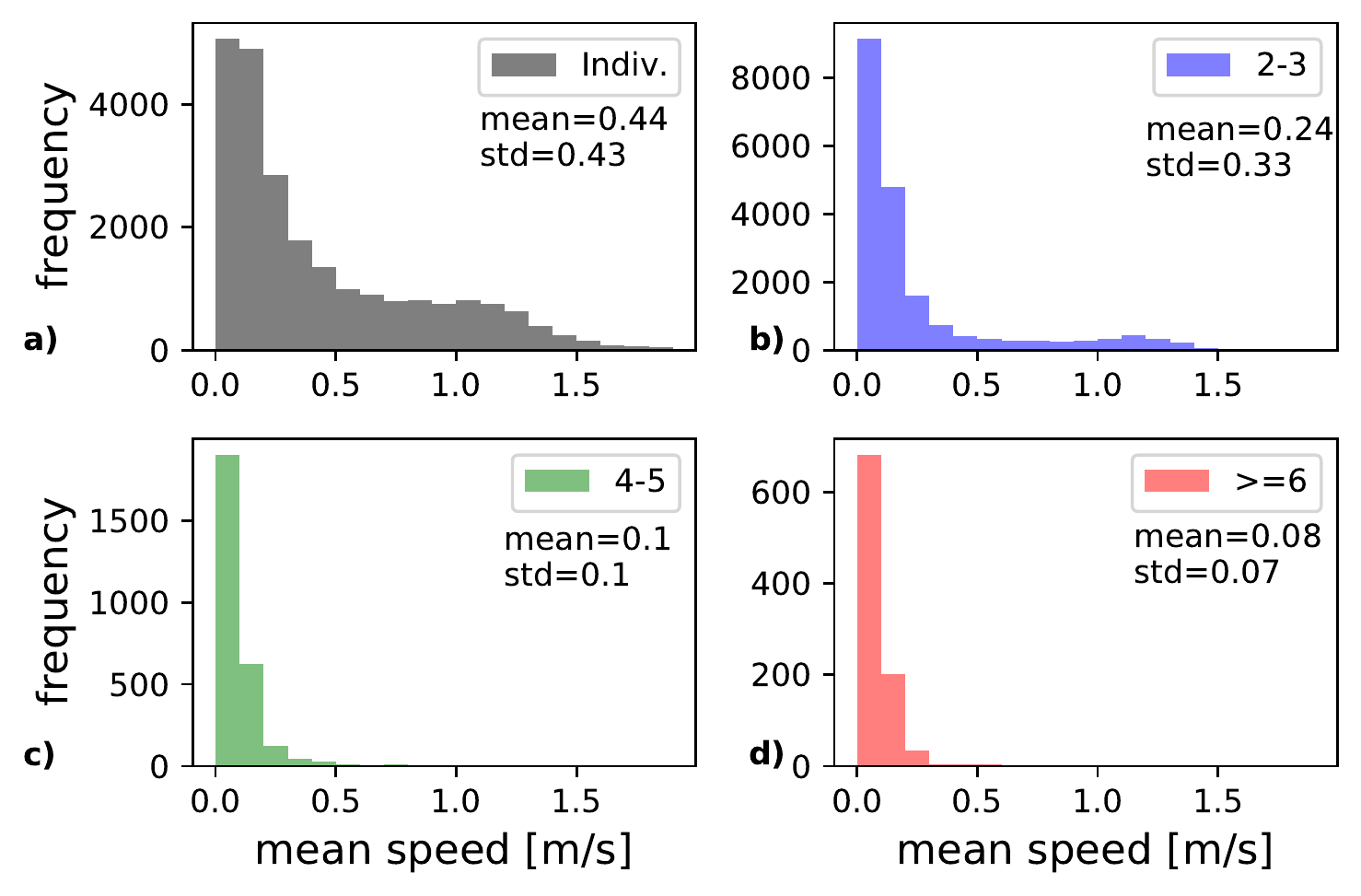}
\caption{ Mean speed of passengers during afternoon peak hours. (a) Individuals (boarding only) (b) Members of pairs and trios c) Groups with 4-5 members (d) Groups with 6 or more members. With increasing group size, the mean speed decreases. }
\label{fig_5}
\end{figure}

Using Eq~(\ref{eq:velocity}), the instantaneous speed and its mean were calculated for each pedestrian. In order to analyse the differences in mean speed distribution of group members and individuals, histograms of the mean speed are shown in Fig~\ref{fig_5} for (a) individuals, (b) groups of two to three members, (c) groups with four to five members and (d) groups with six or more members. The presence of different types of users, namely boarding and alighting passengers, causes the distribution of mean speed to differ significantly between boarding and alighting \cite{kuepper2020}. It is noted that only boarding individual are considered in Fig~\ref{fig_5}a. 
While the average mean speed is 0.24 m/s with a standard deviation of 0.33 for members of pairs and trios, the mean value for groups of 4-5 members is 0.1 m/s with a standard deviation of 0.1. For groups with six or more members the speed is only 0.08 m/s with a standard deviation of 0.07. Comparing the histograms, it becomes apparent that the mean speed decreases with increasing group sizes.

\begin{figure}[htbp]
\includegraphics[width=\textwidth]{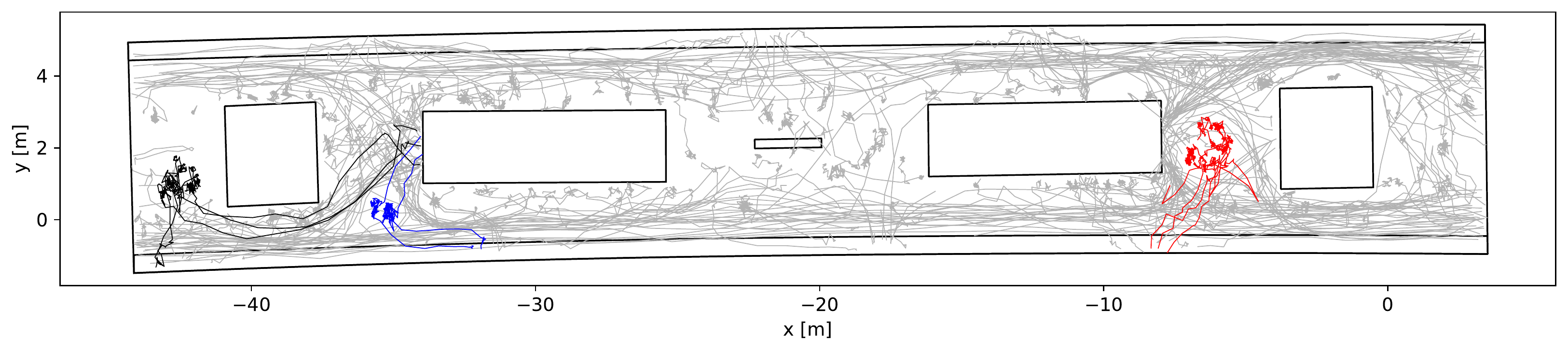}
\caption{ Trajectories of pedestrians at the platform in an exemplary six-minute interval. Trajectories of groups are indicated in colour. Group of two in blue, group of three in black and group of five in red. }
\label{fig_6}
\end{figure}

Fig~\ref{fig_6} illustrates the trajectories of pedestrians in an exemplary six-minute interval. Trajectories of individuals are marked in grey, while the three groups present on the platform in this time interval are highlighted in colour: the group of two in blue, of three in black and the five-member group in red. Due to the chosen time interval, pedestrians' trajectories do not necessarily cover their complete journey through the platform. 
For example, the two groups with two (blue) and three members (black) enter and leave the platform in the considered time interval, causing their trajectories to begin at the stairs and end at a train door. The group with five members (red), on the other hand, was already located at the waiting spot in front of the entrance at the beginning of the selected time interval. Therefore, the trajectories show their waiting positions and the way towards the train, but not the path they chose to enter the platform. 
Waiting passengers close to obstacles influence moving pedestrians to walk closer to the platform edges. Therefore, walking ways can be identified as an accumulation of trajectories in the regions in the vicinity to the safety line.
The trajectories of individuals (grey) indicate the detours in the regions where social groups are waiting. This is especially prominent with the five-person group, which is waiting in front of the entrance at the right-hand side. Similar detours are observable with the waiting pair (blue). However, due to the lesser space requirements of smaller groups, the impact is smaller. The group of three (black) is waiting at the rearward side of the elevator and therefore seems to have no significant influence on travel paths.  

\begin{figure}[tbp]
\centering
\includegraphics[width=0.96\textwidth]{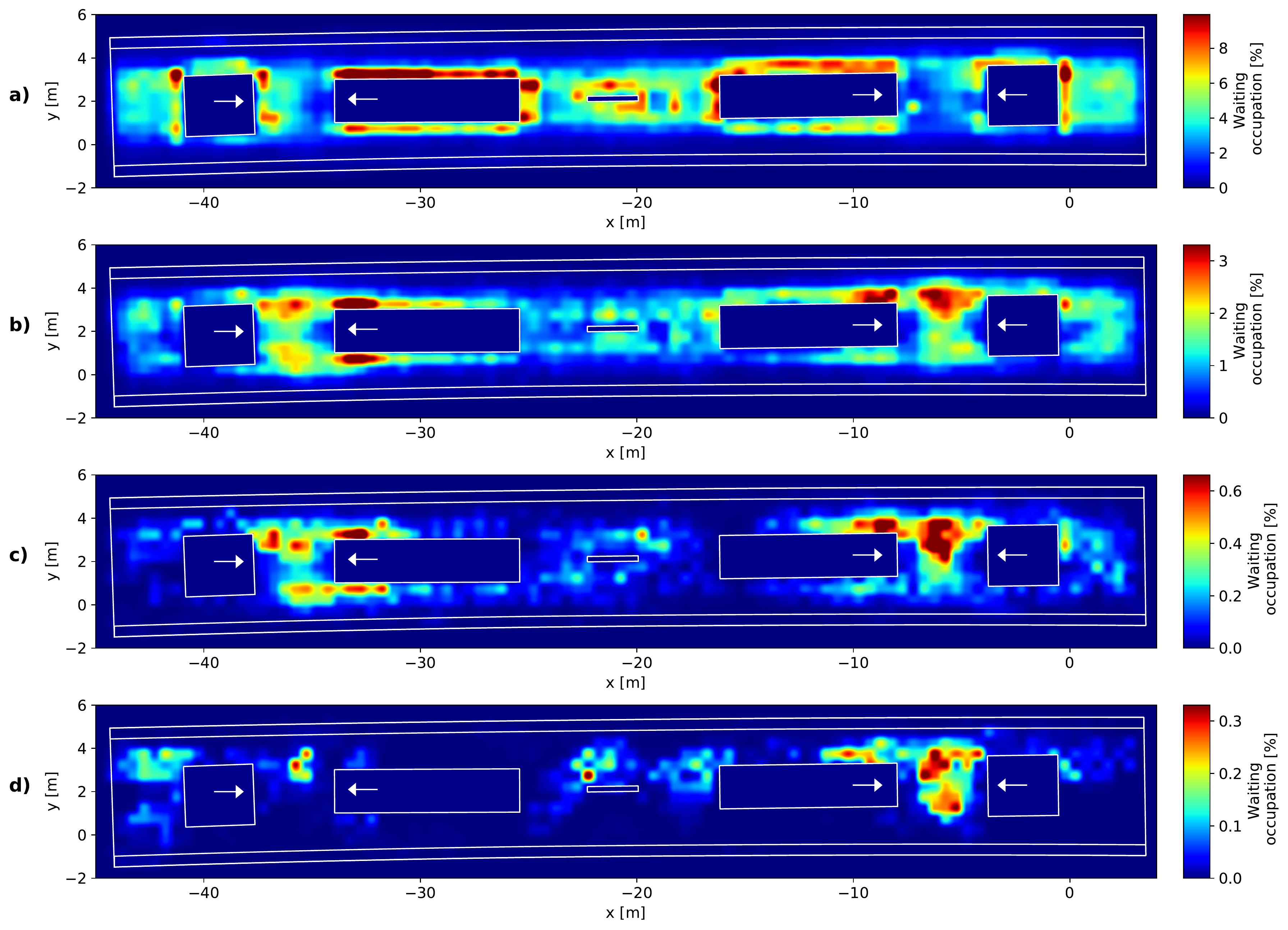}
\caption{ Waiting places of social groups and individuals. (a) individuals (b) groups with 2-3 members (c) groups with 4-5 members (d) groups with 6 or more members. Coloured areas indicate the occupation by waiting passengers as percentage of the total observation time. }
\label{fig_7}
\end{figure}

In order to spatially determine differences in the choice of waiting places, a comparison of waiting places of individuals and groups is illustrated in Fig~\ref{fig_7}. Here, the places where passengers belonging to each group size exhibit a speed below the threshold for waiting were mapped. 
For this purpose, the measurement area was divided into tiles with a size of 0.5 by 0.5 metres, in accordance to the average human shoulder widths \cite{Buchmueller2006}. 
The waiting occupation $O_{wait}$ of a tile with the length $\Delta x$ and the midpoint $\Vec{x}_0$ can be calculated as
\begin{equation}
    O_{wait}= \frac{1}{N_{f}} \sum_{f=0}^{N_f} \sum_i^N \int\displaylimits_{\Vec{x}_0-\Delta \Vec{x}/2 }^{\Vec{x}_0+\Delta \Vec{x}/2 }  \delta( \Vec{x}_{i,f}- \Vec{x})d\Vec{x}
\end{equation}
with $\Vec{x}_{i,l}$ the position of a waiting pedestrian $i$ at the time $f$, $N_f$ the number of frames, $N$ the number of pedestrians and $\delta(x)$ the Dirac delta function. In words: If a pedestrian assigned to the specific group size is considered as waiting in a certain frame, the value attributed to the tile which the pedestrian is occupying is increased by one. By dividing the resulting values by the total observation time (in this study 3 hours on 28 days, which corresponds to 3024000 frames), the measure gives the percentage of time each tile is occupied by a waiting pedestrian.
The direction in which a pedestrian enters the platform from stairs or the elevators is marked with white arrows.
Individuals travelling via Zürich Hardbrücke railway station often choose waiting places close to the sides of stairways or elevators and even wait at the places between the rearward sides of the two stairways in Zürich Hardbrücke (cf. Fig~\ref{fig_7}a, between x= -15 m and x= -25 m). 
Groups of two and three show a similar choice of waiting spots, often tending to wait closer to the platform entries (Fig~\ref{fig_7}b). With increasing group sizes, the preferred waiting places are often chosen in direct vicinity to the entry ways (see Fig~\ref{fig_7}c and d), while the rearward sides of stairs are only seldom used. 
Possible reasons for the lower mean speed in larger groups and the tendency to choose waiting places in front of the entries are difficulties in agreements for waiting place optimisations. Social groups choose waiting places that ensure communication between the group members. Therefore, larger groups tend to form circles, which guarantee eye contact between the members. Comparable to the movement of groups in lines perpendicular to the walking direction, cf. \cite{moussaid2010, schultz2014,zanlungo2014}, this leads to higher space requirements. While the area in front of the entry ways at Zurich Hardbrücke is wide enough for a larger group, the way to the less frequented rearward sides of the stair ways necessitates the passing of the narrower parts of the platform. In order to change the waiting place of a group from the entry ways towards less crowded and frequently used places, an active agreement within and a coordination of the group members is needed. 

\begin{figure}[tbp]
\centering
\includegraphics[width=0.96\textwidth]{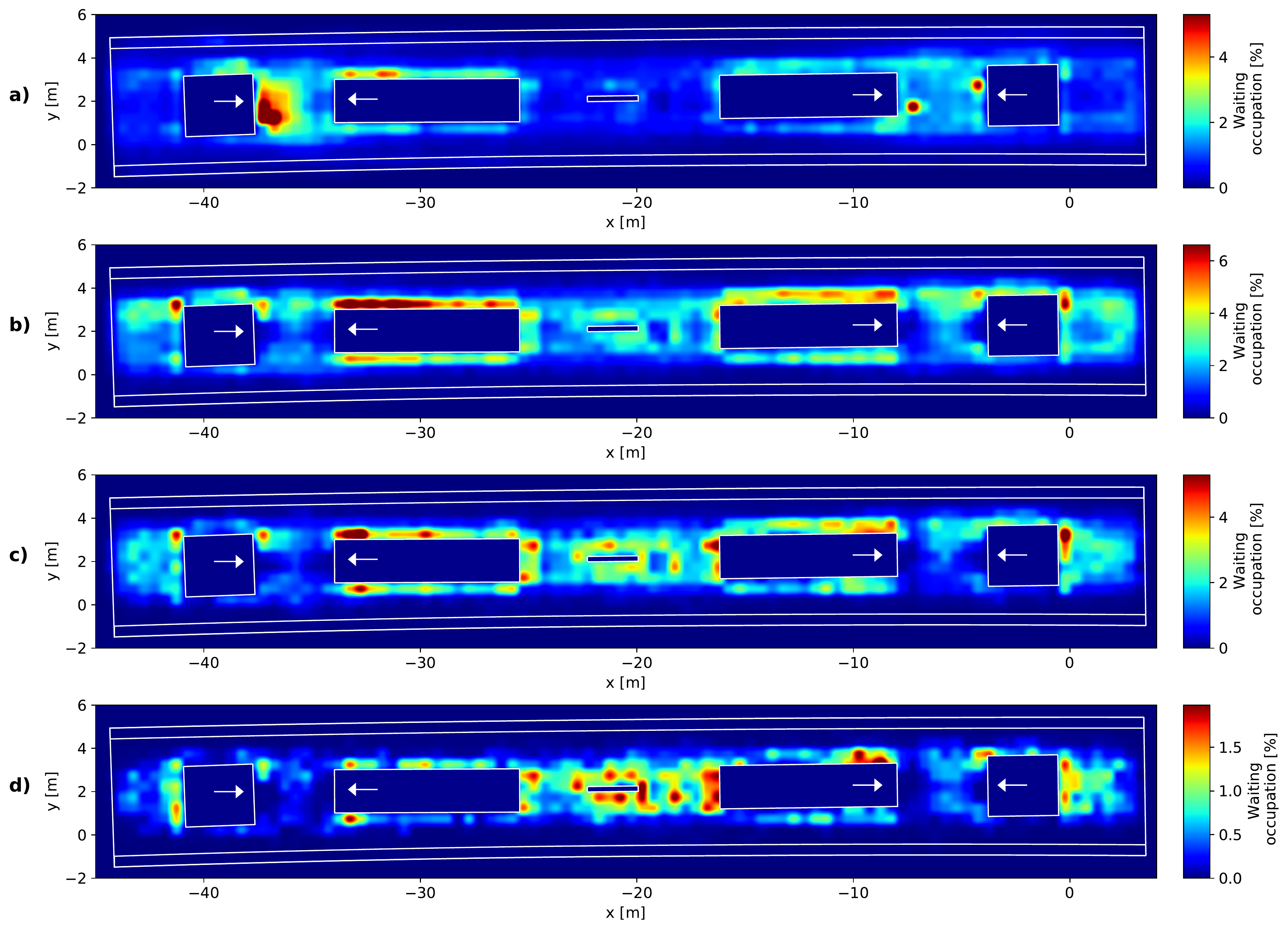}
\caption{ Waiting places of passengers depending on total waiting time. Waiting places of passengers that wait a) up to 2 minutes b) 2-5 minutes c) 5-10 minutes d) 10 and more minutes. Pedestrians with short or intermediate waiting times chose waiting places at the sides of stairs and close to obstacles, with increasing waiting times the rearward sides of stairs are used more frequently.}
\label{fig_8}
\end{figure}

Since the group size is not necessarily the only factor that influences the choice of waiting places, in the following it is analysed how the waiting position is related to the waiting time. The total waiting time for each pedestrian is calculated as the sum of all frames in which the criterion $v_i(t)<0.4 m/s$ is meet. Similar to Fig~\ref{fig_7}, the waiting positions of pedestrians were mapped on a grid with tiles with a size of 0.5 by 0.5 metres and the portion of the total observation time is given.
The results of the mapping are presented in Fig~\ref{fig_8}, these are given regardless of group membership but with respect to the total waiting time of passengers.
Passengers waiting up to 2 minutes chose locations close to the entry ways (Fig~\ref{fig_8}a), as those are meant for passengers that arrive only shortly before the train they intend to board.  
With increasing waiting time (2-5 minutes in Fig~\ref{fig_8}b and 5-10 minutes in Fig~\ref{fig_8}c) obstacles with the possibility to lean against, for example the sides of the stairs or the information boards, are used for waiting.
Passengers with long waiting times (10 minutes and longer), (Fig~\ref{fig_8}d) mainly chose waiting places in rearward areas behind the elevators ($x >~0~m$ or $x < -40~m$) or between the two stair ways ($-25 ~m <x< -15 ~m$). These are the places that are least frequently used by walking pedestrians and therefore the least disturbed. Passengers tend to wait in places that allow to turn in the direction of the next intended action. In most cases in the context of railway platforms this is the boarding process.

Pedestrians waiting in front of the entrances are likely to interfere with the passenger flow at the platform, since they narrow the available space for passing pedestrians in an area where the most movements occur. As shown in both Fig~\ref{fig_7} and Fig~\ref{fig_8} this is likely the case for passengers that are members of larger social groups or that only wait for a short time. This information can be used to optimise the distribution of waiting passengers at railway platforms. Passengers with short waiting time arrive with little head time to the trains and will therefore choose waiting spots close to the entrances. However, social groups likely wait in those areas because of their higher space requirements. The planning of infrastructure modifications would thus do well to consider the requirements of social groups in order to achieve even passenger distributions and ensure the flow at the platform entrances. This is also important for the assessment of comfort of waiting passengers by e.g. level of service concepts. 

The discussion of the results of the previous sections leads to the subsequent findings: Passengers tend to choose their waiting places in accordance with the following criteria: (a) short walking distances, (b) possibility to turn towards the next intended action (most likely the boarding process), (c) being undisturbed by other passengers or avoiding disturbing others and  (d) ensuring communication. These are very similar to the criteria determined in the context of an inflow in a confined space (like an elevator) \cite{ezaki2016}, where flow avoidance, distance cost and boundary preferences were suggested.
The ranking of these criteria and therewith the assessment of comfort differs with varying types of passengers.
Passengers with short waiting times choose places with short walking distances close to the expected stopping position of the trains. With longer waiting times the criterion of undisturbed waiting places becomes more important and out-weights the preference of the shortest distances. 
However, in contrast to individuals, communication is the dominant criterion for social groups. Therefore, social groups do not necessarily wait in places where they do not disturb others. In order to ensure eye contact and communication to all group members, waiting social groups form circles and therefore have higher space requirements, leading them to act as obstacles for passing pedestrians.

\section{Conclusion}
This paper presented a method that allows the identification of social groups in trajectory data of waiting and standing pedestrians.
Social groups were identified by thresholds for inter-personal distances that are present over certain time intervals. In the case of a train station platform, two pedestrians are considered as belonging to a social group, if their distance to one another is smaller than 1.5 m for 85 \% of the time in which they are simultaneously inside the observation area and smaller than 1 m for 40 \% of that time. 
The percentages of the time that are needed for pedestrians to be considered as a social group were determined by a parameter study and were checked against a ground truth for validation. The ground truth was established by a visual analysis performed by two independent test persons. However, these parameters need to be reconsidered and validated for different scenarios, e.g. in shopping malls or public gatherings where different dynamics are occurring. The group detection is suitable for scenarios with low densities; the applicability in dense environments cannot be guaranteed as in those cases small distances between pedestrians are caused by congestion or limited available space. 

The group detection method was applied to a data set taken from the afternoon peak hours during February 2020 in Zürich Hardbrücke, Switzerland. 
During working days about 9-10 \% of the pedestrians waiting at the train station platform were members of social groups; the portion increases to up to 20 \% during weekends.
The most frequently observed group size was pairs, each size is less frequent than the next smaller size. Distributions of group sizes showed no correlation to whether it was a working day or weekend day. With increasing group size, the members mean speed decreased.
While individuals often waited at the sides of stairs and elevators, social groups were found to be more likely to choose waiting places that provide enough space for members to position themselves in such a way that enables communication within the group. Typically, this is the case in the vicinity of the platform entrances. This behaviour was shown to be more prominent with increasing group size. 
Moreover, waiting places were influenced by the total waiting time of the passengers. Pedestrians with short waiting times (less than 2 minutes) waited close the entrances. For longer waiting times places at the undisturbed rearward sides of the stairs were used.

The waiting places chosen by individuals and groups highlight the different needs in terms of comfort.
The waiting places were chosen based on a ranking of the criteria of short walking distances, the direction of the train arrival, undisturbed waiting places and ensured communication. Depending on the types of users and the waiting time those criteria were prioritised differently. 
Passengers with long waiting times prefer undisturbed waiting places even if the distance was longer. While individuals chose undisturbed waiting places in areas where they do not hinder the movement of others, social groups prioritised the possibility to communicate even if the position was close to the highly frequented entry way. The results could be used to assess the comfort of different types of users by level of service concept including waiting passengers and to optimise space usage at railway platforms by increasing the robustness of performance during peak load by optimising the pedestrian distribution. 

\section*{Acknowledgments}
The authors would like to thank the Swiss Federal Railways for providing the data and the student assistants for their help with the visual group assessment.

\section*{Author Contributions}{Conceptualization: M.K. and A.S.;  Data curation: M.K.; Formal analysis: M.K.; Funding acquisition: A.S.; Investigation: M.K.; Methodology: M.K. and A.S.; Project administration: M.K. and A.S.; Software: M.K.; Supervision: A.S.; Validation: M.K.; Visualization: M.K.; Writing--original draft preparation: M.K.; Writing--review and editing: M.K. and A.S.}

\section*{Funding} The study was part of the project “CroMa — Crowd Management in Verkehrsinfrastrukturen.” The project was funded by the German Federal Ministry of Education and Research under grant number 13N14530. The funders had no role in the design of the study; in the collection, analyses or interpretation of data; in the writing of the manuscript, or in the decision to publish the results.


\begin{thebibliography}{999}
\providecommand{\natexlab}[1]{#1}

\bibitem{predtetschenski1971}
Predtetschenski WM, Milinski AI.
\newblock Personenstr{\"o}me in {G}eb{\"a}uden - {B}erechnungsmethoden f{\"u}r
  die {P}rojektierung.
\newblock Verlagsgesellschaft Rudolf M{\"u}ller. 1971;.

\bibitem{boltes2018}
Boltes M, Zhang J, Tordeux A, Schadschneider A, Seyfried A.
\newblock Empirical results of pedestrian and evacuation dynamics.
\newblock Encyclopedia of complexity and systems science. 2018; p. 1--29.

\bibitem{chraibi2018}
Chraibi M, Tordeux A, Schadschneider A, Seyfried A.
\newblock Modelling of pedestrian and evacuation dynamics.
\newblock Encyclopedia of Complexity and Systems Science. 2018; p. 1--22.

\bibitem{ped2018}
Dederichs A, K{\"o}ster G, Schadschneider A.
\newblock Proceedings of Pedestrian and Evacuation Dynamics 2018.
\newblock Collective Dynamics. 2020;5:1--543.

\bibitem{tgf2020}
Zuriguel I, Garcimart{\'\i}n A, Hidalgo RC.
\newblock Traffic and Granular Flow 2019.
\newblock Springer; 2020.

\bibitem{Weidmann1993}
Weidmann U.
\newblock Transporttechnik der {F}u{\ss}g{\"a}nger: transporttechnische
  {E}igenschaften des {F}u{\ss}g{\"a}ngerverkehrs, {L}iteraturauswertung.
\newblock IVT Schriftenreihe. 1993;90.

\bibitem{Buchmueller2006}
Buchmueller S, Weidmann U.
\newblock Parameters of pedestrians, pedestrian traffic and walking facilities.
\newblock IVT Schriftenreihe. 2006;132.

\bibitem{Fruin1971}
Fruin JJ.
\newblock Pedestrian planning and design.
\newblock Elevator World, Inc, Mobile, AL. 1971;.

\bibitem{Pettersson2011}
Pettersson P. Passenger waiting strategies on railway platforms-Effects of
  information and platform facilities-: Case study Sweden and Japan; 2011.

\bibitem{zhou2020}
Zhou M, Ge S, Liu J, Dong H, Wang FY.
\newblock Field observation and analysis of waiting passengers at subway
  platform—A case study of Beijing subway stations.
\newblock Physica A: Statistical Mechanics and its Applications.
  2020;556:124779.

\bibitem{zhou2019}
Zhou M, Dong H, Wang FY, Zhao Y, Gao S, Ning B.
\newblock Field observations and modeling of waiting pedestrian at subway
  platform.
\newblock Information Sciences. 2019;504:136--160.

\bibitem{krstanoski2014}
Krstanoski N.
\newblock MODELLING PASSENGER DISTRIBUTION ON METRO STATION PLATFORM.
\newblock International Journal for Traffic \& Transport Engineering.
  2014;4(4).

\bibitem{Bosina2015}
Bosina E, Britschgi S, Meeder M, Weidmann U.
\newblock Distribution of passengers on railway platforms.
\newblock In: Proceedings of the 15th Swiss Transport Research Conference;
  2015.

\bibitem{Davidich2013}
Davidich M, K{\"o}ster G.
\newblock Predicting pedestrian flow: A methodology and a proof of concept
  based on real-life data.
\newblock PloS one. 2013;8(12):e83355.

\bibitem{Lam1999}
Lam WH, Cheung CY, Lam C.
\newblock A study of crowding effects at the Hong Kong light rail transit
  stations.
\newblock Transportation Research Part A: Policy and Practice.
  1999;33(5):401--415.

\bibitem{ingvardson2018}
Ingvardson JB, Nielsen OA, Raveau S, Nielsen BF.
\newblock Passenger arrival and waiting time distributions dependent on train
  service frequency and station characteristics: A smart card data analysis.
\newblock Transportation Research Part C: Emerging Technologies.
  2018;90:292--306.

\bibitem{lee2007}
Lee Yc, Daamen W, Wiggenraad P.
\newblock Boarding and alighting behavior of public transport passengers.
\newblock 86th Annual Meeting Transportation Research Board. 2007; p. 1--14.

\bibitem{Bosina2017b}
Bosina E, Meeder M, Weidmann U.
\newblock Pedestrian flows on railway platforms.
\newblock In: Swiss Transport Research Conference. vol.~17; 2017.

\bibitem{turner1987}
Turner JC.
\newblock The analysis of social influence.
\newblock Rediscovering the social group: A self-categorization theory. 1987;
  p. 68--88.

\bibitem{drury2020}
Drury J, Reicher S.
\newblock Crowds and collective behavior.
\newblock In: Oxford research encyclopedia of psychology; 2020.

\bibitem{templeton2020}
Templeton A, Drury J, Philippides A.
\newblock Placing large group relations into pedestrian dynamics: Psychological
  crowds in counterflow.
\newblock Collective Dynamics. 2020;4:1--22.

\bibitem{tajfel1978}
Tajfel HE.
\newblock Differentiation between social groups: Studies in the social
  psychology of intergroup relations.
\newblock Academic Press; 1978.

\bibitem{turner2010}
Turner JC.
\newblock Towards a cognitive redefinition of the social group.
\newblock In: Research Colloquium on Social Identity of the European Laboratory
  of Social Psychology, Dec, 1978, Universit{\'e} de Haute Bretagne, Rennes,
  France; This chapter is a revised version of a paper first presented at the
  aforementioned colloquium. Psychology Press; 2010.

\bibitem{moussaid2010}
Moussa{\"\i}d M, Perozo N, Garnier S, Helbing D, Theraulaz G.
\newblock The walking behaviour of pedestrian social groups and its impact on
  crowd dynamics.
\newblock PloS one. 2010;5(4):e10047.

\bibitem{schultz2014}
Schultz M, R{\"o}{\ss}ger L, Fricke H, Schlag B.
\newblock Group dynamic behavior and psychometric profiles as substantial
  driver for pedestrian dynamics.
\newblock Pedestrian and evacuation dynamics 2012. 2014; p. 1097--1111.

\bibitem{zanlungo2014}
Zanlungo F, Ikeda T, Kanda T.
\newblock Potential for the dynamics of pedestrians in a socially interacting
  group.
\newblock Physical Review E. 2014;89(1):012811.

\bibitem{zanlungo2014b}
Zanlungo F, Br{\v{s}}{\v{c}}i{\'c} D, Kanda T.
\newblock Pedestrian group behaviour analysis under different density
  conditions.
\newblock Transportation Research Procedia. 2014;2:149--158.

\bibitem{zanlungo2015}
Zanlungo F, Br{\v{s}}{\v{c}}i{\'c} D, Kanda T.
\newblock Spatial-size scaling of pedestrian groups under growing density
  conditions.
\newblock Physical Review E. 2015;91(6):062810.

\bibitem{james1953}
James J.
\newblock The distribution of free-forming small group size.
\newblock American Sociological Review. 1953;.

\bibitem{coleman1961}
Coleman JS, James J.
\newblock The equilibrium size distribution of freely-forming groups.
\newblock Sociometry. 1961;24(1):36--45.

\bibitem{pouw2021}
Pouw CA, Willems J, van Schadewijk F, Thurau J, Toschi F, Corbetta A.
\newblock Benchmarking high-fidelity pedestrian tracking systems for research,
  real-time monitoring and crowd control.
\newblock Collective Dynamics 6 (2021): 1-22.

\bibitem{corbetta2014}
Corbetta A, Bruno L, Muntean A, Toschi F.
\newblock High statistics measurements of pedestrian dynamics.
\newblock Transportation Research Procedia. 2014;2:96--104.

\bibitem{yucel2013}
Y{\"u}cel Z, Zanlungo F, Ikeda T, Miyashita T, Hagita N.
\newblock Deciphering the crowd: Modeling and identification of pedestrian
  group motion.
\newblock Sensors. 2013;13(1):875--897.

\bibitem{yucel2019}
Yucel Z, Zanlungo F, Feliciani C, Gregorj A, Kanda T.
\newblock Identification of social relation within pedestrian dyads.
\newblock PloS one. 2019;14(10):e0223656.

\bibitem{brscic2014}
Br{\v{s}}{\v{c}}i{\'c} D, Zanlungo F, Kanda T.
\newblock Density and velocity patterns during one year of pedestrian tracking.
\newblock Transportation Research Procedia. 2014;2:77--86.

\bibitem{pouw2020}
Pouw CA, Toschi F, van Schadewijk F, Corbetta A.
\newblock Monitoring physical distancing for crowd management: Real-time
  trajectory and group analysis.
\newblock PloS one. 2020;15(10):e0240963.

\bibitem{aveni1977}
Aveni AF.
\newblock The not-so-lonely crowd: Friendship groups in collective behavior.
\newblock Sociometry. 1977; p. 96--99.

\bibitem{reuter2012}
Reuter V, Bergner BS, K{\"o}ster G, Seitz M, Treml F, Hartmann D.
\newblock On Modeling Groups in Crowds: Empirical Evidence and Simulation
  Results Including Large Groups.
\newblock In: Weidmann U, Kirsch U, Schreckenberg M, editors. Pedestrian and
  Evacuation Dynamics 2012. Cham: Springer International Publishing; 2014. p.
  835--845.

\bibitem{templeton2018}
Templeton A, Drury J, Philippides A.
\newblock Walking together: behavioural signatures of psychological crowds.
\newblock Royal Society open science. 2018;5(7):180172.

\bibitem{vonKruechten2017}
Von~Kr{\"u}chten C, Schadschneider A.
\newblock Empirical study on social groups in pedestrian evacuation dynamics.
\newblock Physica A: Statistical Mechanics and its Applications.
  2017;475:129--141.

\bibitem{bode2015}
Bode NW, Holl S, Mehner W, Seyfried A.
\newblock Disentangling the impact of social groups on response times and
  movement dynamics in evacuations.
\newblock PloS one. 2015;10(3):e0121227.

\bibitem{vanHeuvel2017}
van~den Heuvel J, Thurau J, Mendelin M, Schakenbos R, van Ofwegen M,
  Hoogendoorn SP.
\newblock An application of new pedestrian tracking sensors for evaluating
  platform safety risks at Swiss and Dutch train stations.
\newblock In: International Conference on Traffic and Granular Flow. Springer;
  2017. p. 277--286.

\bibitem{lohner2010}
L{\"o}hner R.
\newblock On the modeling of pedestrian motion.
\newblock Applied Mathematical Modelling. 2010;34(2):366--382.

\bibitem{attali2003}
Attali D, Boissonnat JD, Lieutier A.
\newblock Complexity of the delaunay triangulation of points on surfaces the
  smooth case.
\newblock In: Proceedings of the nineteenth annual symposium on Computational
  Geometry; 2003. p. 201--210.

\bibitem{abellanas1999}
Abellanas M, Hurtado F, Ramos PA.
\newblock Structural tolerance and Delaunay triangulation.
\newblock Information Processing Letters. 1999;71(5-6):221--227.

\bibitem{kuepper2020}
K{\"u}pper M, Seyfried A.
\newblock Analysis of Space Usage on train station platforms based on
  trajectory data.
\newblock Sustainability. 2020;12(20):8325.

\bibitem{Jupedsim2022}
Team DC. Ju{P}ed{S}im; 2022.
\newblock https://doi.org/10.5281/zenodo.6144559.

\bibitem{ezaki2016}
Ezaki T, Ohtsuka K, Chraibi M, Boltes M, Yanagisawa D, Seyfried A, et~al.
\newblock Inflow Process of Pedestrians to a Confined Space.
\newblock Collective Dynamics. 2016;1:1–18.
\newblock doi:{10.17815/CD.2016.4}.

\end{thebibliography}
\end{document}